\documentclass[reprint, superscriptaddress, prl, floatfix]{revtex4-2}
\usepackage{amsmath,amssymb}
\usepackage{graphicx}
\usepackage{bm}
\usepackage{times}
\usepackage{xcolor}
\usepackage{lineno}
\usepackage{textcomp, gensymb}
\usepackage{hyperref}
\usepackage{tabularx}
\usepackage{adjustbox}

\newcommand{\nuebar}{\ensuremath{\overline{\nu}_e}}
\newcommand{\uFive}{$^{235}$U}

\newcommand{\DXNull}{13.6}
\newcommand{\DXAll}{3.9}
\newcommand{\DXIso}{0.4}
\newcommand{\CLNull}{3.7}
\newcommand{\CLAll}{2.0}

\newcommand{\NIST}{National Institute of Standards and Technology (NIST) \renewcommand{\NIST}{NIST}}
\newcommand{\NBSR}{National Bureau of Standards Reactor (NBSR) \renewcommand{\NBSR}{NBSR}}
\newcommand{\INL}{Idaho National Laboratory (INL) \renewcommand{\INL}{INL}}
\newcommand{\ATR}{Advanced Test Reactor (ATR) \renewcommand{\ATR}{ATR}}
\newcommand{\ORNL}{Oak Ridge National Laboratory (ORNL) \renewcommand{\ORNL}{ORNL}}
\newcommand{\HFIR}{High Flux Isotope Reactor (HFIR) \renewcommand{\HFIR}{HFIR}}
\newcommand{\LLNL}{Lawrence Livermore National Laboratory (LLNL) \renewcommand{\LLNL}{LLNL}}

\usepackage{verbatim}

\begin{document}


\title{Final Measurement of the \uFive{} Antineutrino Energy Spectrum with the PROSPECT-I Detector at HFIR}

\address{Department of Physics, Boston University, Boston, MA, USA} \vspace{-0.4\baselineskip}
\address{Brookhaven National Laboratory, Upton, NY, USA} \vspace{-0.4\baselineskip}
\address{Department of Physics, Drexel University, Philadelphia, PA, USA} \vspace{-0.4\baselineskip}
\address{George W.\,Woodruff School of Mechanical Engineering, Georgia Institute of Technology, Atlanta, GA, USA} \vspace{-0.4\baselineskip}
\address{Department of Physics and Astronomy, University of Hawaii, Honolulu, HI, USA} \vspace{-0.4\baselineskip}
\address{Department of Physics, Illinois Institute of Technology, Chicago, IL, US} \vspace{-0.4\baselineskip}
\address{Nuclear and Chemical Sciences Division, Lawrence Livermore National Laboratory, Livermore, CA, USA} \vspace{-0.4\baselineskip}
\address{Department of Physics, Le Moyne College, Syracuse, NY, USA} \vspace{-0.4\baselineskip}
\address{National Institute of Standards and Technology, Gaithersburg, MD, USA} \vspace{-0.4\baselineskip}
\address{High Flux Isotope Reactor, Oak Ridge National Laboratory, Oak Ridge, TN, USA} \vspace{-0.4\baselineskip}
\address{Physics Division, Oak Ridge National Laboratory, Oak Ridge, TN, USA} \vspace{-0.4\baselineskip}
\address{Department of Physics, Susquehanna University, Selinsgrove, PA, USA} \vspace{-0.4\baselineskip}
\address{Department of Physics, Temple University, Philadelphia, PA, USA} \vspace{-0.4\baselineskip}
\address{Department of Physics and Astronomy, University of Tennessee, Knoxville, TN, USA} \vspace{-0.4\baselineskip}
\address{Department of Physics, United States Naval Academy, Annapolis, MD, USA} \vspace{-0.4\baselineskip}
\address{Institute for Quantum Computing and Department of Physics, University of Waterloo, Waterloo, ON, Canada} \vspace{-0.4\baselineskip}
\address{Department of Physics, University of Wisconsin, Madison, WI, USA} \vspace{-0.4\baselineskip}
\address{Wright Laboratory, Department of Physics, Yale University, New Haven, CT, USA} \vspace{-0.4\baselineskip}

\author{M.~Andriamirado$^{6}$,
A.~B.~Balantekin$^{17}$,
C.~D.~Bass$^{8}$,
D.~E.~Bergeron$^{9}$,
E.~P.~Bernard$^{7}$,
N.~S.~Bowden$^{7}$,
C.~D.~Bryan$^{10}$,
R.~Carr$^{ 15}$, 
T.~Classen$^{7}$,
A.~J.~Conant$^{10}$,
G.~Deichert$^{10}$,
A.~Delgado$^{11}$,
M.~V.~Diwan$^{2}$,
M.~J.~Dolinski$^{3}$,
A.~Erickson$^{4}$,
B.~T.~Foust$^{18}$,
J.~K.~Gaison$^{18}$,
A.~Galindo-Uribari$^{11,14}$,
C.~E.~Gilbert$^{11}$,
S.~Gokhale$^{2}$,
C.~Grant$^{1}$, 
S.~Hans$^{2}$,
A.~B.~Hansell$^{12}$,
K.~M.~Heeger$^{18}$,
B.~Heffron$^{11,14}$,
D.~E.~Jaffe$^{2}$,
S.~Jayakumar$^{3}$,
X.~Ji$^{2}$,
D.~C.~Jones$^{13}$, 
J.~Koblanski$^{5}$,
P.~Kunkle$^{1}$, 
O.~Kyzylova$^{3}$, 
D.~LaBelle$^{3}$,
C.~E.~Lane$^{3}$,
T.~J.~Langford$^{18}$,
J.~LaRosa$^{9}$, 
B.~R.~Littlejohn$^{6}$,
X.~Lu$^{11,14}$, 
J.~Maricic$^{5}$,
M.~P.~Mendenhall$^{7}$,
A.~M.~Meyer$^{5}$, 
R.~Milincic$^{5}$, 
P.~E.~Mueller$^{11}$,
H.~P.~Mumm$^{9}$,
J.~Napolitano$^{13}$,
R.~Neilson$^{3}$,
J.~A.~Nikkel$^{18}$, 
S.~Nour$^{9}$, 
J.~L.~Palomino Gallo$^{6}$, 
D.~A.~Pushin$^{16}$,
X.~Qian$^{2}$,
C.~Roca$^{7}$,
R.~Rosero$^{2}$,
M.~Searles$^{10}$,
P.~T.~Surukuchi$^{18}$,
F.~Sutanto$^{7}$,
M.~A.~Tyra$^{9}$, 
D.~Venegas-Vargas$^{11,14}$, 
P.~B.~Weatherly$^{3}$, 
J.~Wilhelmi$^{18}$,
A.~Woolverton$^{16}$, 
M.~Yeh$^{2}$,
C.~Zhang$^{2}$ and
X.~Zhang$^{7}$ \\
}


\collaboration{The PROSPECT Collaboration}
\homepage{http://prospect.yale.edu \\ email: prospect.collaboration@gmail.com}

\begin{abstract}
This Letter reports one of the most precise measurements to date of the antineutrino spectrum from a purely \uFive{}-fueled reactor, made with the final dataset from the PROSPECT-I detector at the High Flux Isotope Reactor. By extracting information from previously unused detector segments, this analysis effectively doubles the statistics of the previous PROSPECT measurement. The reconstructed energy spectrum is unfolded into antineutrino energy and compared with both the Huber-Mueller model and a spectrum from a commercial reactor burning multiple fuel isotopes. A local excess over the model is observed in the 5~MeV to 7~MeV energy region. Comparison of the PROSPECT results with those from commercial reactors provides new constraints on the origin of this excess, disfavoring at \CLAll{} and \CLNull{} standard deviations the hypotheses that antineutrinos from \uFive{} are solely responsible and non-contributors to the excess observed at commercial reactors respectively.
\end{abstract}

\maketitle



Nuclear reactors, among the brightest terrestrial emitters of antineutrinos ($\bar{\nu}_e$), have been central to neutrino physics. Since the antineutrino was first observed at a reactor \cite{cowan1956}, increasingly precise experiments have measured the long-baseline flavor mixing driven by $\theta_{12}$ \cite{KamLAND_rate}, revealed shorter-baseline flavor oscillations driven by $\theta_{13}$ ~\cite{db_prl_shape, reno_shape, dc_nature}, and searched for sterile neutrino-driven oscillations at shorter distances \cite{danss_osc,neos,stereo_2019, Neutrino-4,  PROSPECT:2020sxr}. Recently, it has become clear that phenomenological models do not capture the full physics of antineutrino emission from reactor cores. Observed reactor antineutrino energy spectra disagree \cite{DayaBay:bump, DC:bump, RENO:bump, stereo_2019, PROSPECT:2020sxr} with predictions based both on beta-spectrum conversion \cite{hayen_initio,HM:Huber}, and \textit{ab initio} calculations from nuclear databases \cite{fallot2}, reporting an excess in the region between 5-7 MeV. The origin of the disagreement is unknown, as is whether it is present for all fissioning isotopes or is dominated by mismodelling of a single isotope. This distinction is challenging to make at commercial reactors with low-enriched uranium (LEU) cores, which burn a time-evolving mixture of isotopes. Beyond shedding light on the modeling discrepancy, precise measurements of antineutrino spectra may themselves be useful for future reactor-based experiments and reactor antineutrino applications \cite{bib:IAEA, Romano:2022spd}.

The PROSPECT antineutrino detector and experimental location at the High Flux Isotope Reactor (HFIR) are described in~\cite{PROSPECT:2018dnc}. HFIR is a 85~MW$_\mathrm{th}$ compact core research reactor that uses 93~\% enriched \uFive{} fuel (HEU). 
The combination of HEU fuel and full core replacement every 24-day reactor cycle means that fuel evolution is negligible and over 99~\% of emitted \nuebar{} are due to ~\uFive{} fission. The detector is located at surface level with minimal overburden, at an average baseline of 7.9\,m from the reactor core.


The detector comprises a single scintillator tank optically separated into 154 segments (14.5~cm $\times$ 14.5~cm $\times$ 117.6~cm), each readout by two PMTs~\cite{Ashenfelter_2019}. 
Approximately 4 tons of $^6$Li-loaded liquid scintillator (LiLS) with good pulse shape discrimination (PSD) properties are used to reject fast neutron recoil backgrounds and identify neutron captures via the $^6\textrm{Li}(n,t)\alpha$ interaction. Neutrinos are detected via Inverse Beta Decay (IBD), in which an \nuebar{} interacts with a proton in the LiLS, producing a positron and a neutron. IBD events are identified via the spatial-temporal correlation of a prompt electromagnetic deposition ($e^+$ ionization and annihilation) and a delayed neutron capture on $^6$Li ($50~\mu$s mean capture time). 

Intrinsic, external, and cosmogenic radiation sources are used to establish the detector’s energy scale, characterize differences in response between segments, and correct for time variations in detector performance~\cite{PROSPECT:2020sxr, PROSPECT2:Calibration}. A detailed \textsc{Geant4}~\cite{Mahmood:1027671} Monte Carlo model of the detector is tuned to accurately reproduce calibration-derived energy and segment multiplicity distributions from multiple sources spanning a range of energies between $0.5$-$13.4$~MeV. This tuned simulation model is used to predict the response matrix that connects incident \nuebar{} energy to the observed prompt energy.

During operation, a number of PMTs gradually became inoperable because of current instabilities. In addition, there was a gradual decrease in the LiLS light yield. 
Previous PROSPECT analyses~\cite{PROSPECT:2018dtt,PROSPECT:2018snc,PROSPECT:2020sxr} used a single detector configuration excluding all segments with any PMT inoperable at any point during data collection, thereby discarding information from earlier time periods where more PMTs were functional.

In this Letter, we introduce improved event reconstruction and analysis techniques that take advantage of multiple detector configurations to more efficiently use information from segments that were fully and/or partially instrumented for part of the data collection period. The resulting dataset yields a significant increase in statistical power thanks to the increased active-detector volume and improved background rejection.



In PROSPECT waveform analysis, features from individual PMTs are grouped into multi-segment \emph{clusters} within a 20~ns arrival time window. The paired waveforms of PMTs on opposite ends of a double-ended (DE) segment are stored and analyzed together as a reconstructed \emph{pulse} containing segment-level information.
Single-ended (SE) segments with only one operable PMT also have waveform features stored and reconstructed as pulses but the position dependence of scintillation light collection  means that deposited energy cannot be accurately reconstructed~\cite{PROSPECT:2015rce, PROSPECT:2018hzo, stereo_2019, Neutrino-4}. Previous analysis only used information from DE segments, while here Single-Ended Event Reconstruction (SEER) is added. SE pulses were found to produce well-separated  electronic and nuclear recoil PSD distributions across the relevant range of PMT pulse amplitudes, providing a mechanism to suppress background without full deposited energy information.

The SEER PSD parameter is determined from  PMT pulse integrals as described in~\cite{PROSPECT:2020sxr}, providing the mean and width of the electromagnetic and nuclear recoil PSD distributions as a function of SE reconstructed energy (E$_{rec}$) and data collection period for each SE segment. 
SEER-determined event information enters into the IBD candidate selection as follows: 
(1) if a delayed cluster includes any SE pulses, it is rejected, since neutron capture on $^6$Li is localized in a single segment; 
(2) if a prompt cluster contains SE pulses with E$_{rec}<$ 0.8 MeV and a PSD value $3.5$~$\sigma$ above the mean of the electromagnetic PSD distribution, the cluster is rejected, since it likely contains a nuclear recoil; 
(3) if a prompt cluster contains SE pulses with E$_{rec}$ $> 0.8$~MeV , it is rejected regardless of PSD to provide enhanced screening of $\gamma$-like backgrounds; 
and (4) IBD candidates within 170~$\mu$s of a cluster containing only high-PSD SE segments are vetoed. 

\begin{figure}[b]
  \centering
  \includegraphics[width=0.48\textwidth]{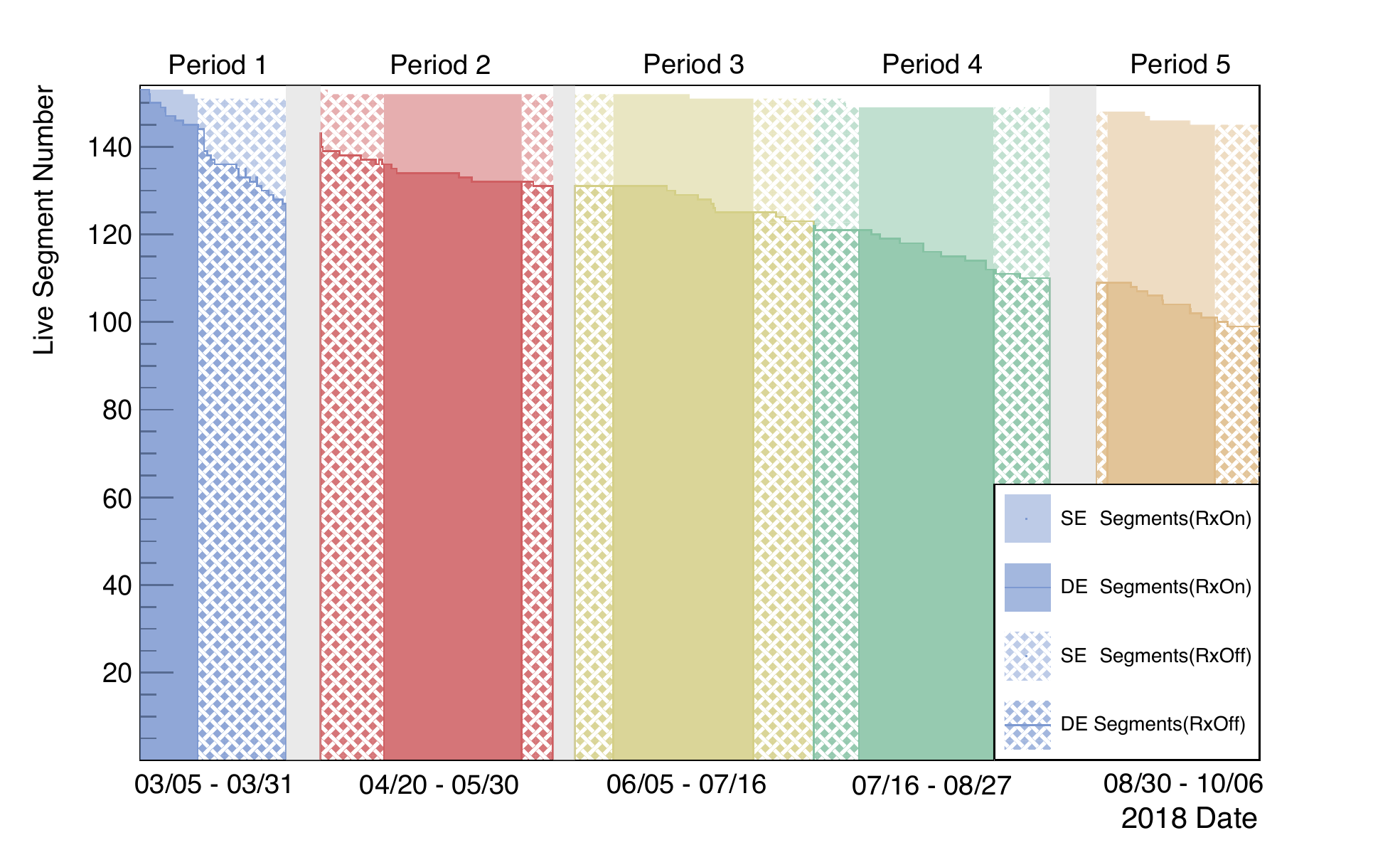}
  \caption{Time evolution of DE and SE segment numbers with reactor-on (RxOn) and reactor-off(RxOff) running, and definition of the 5 DS periods. The number of DE segments can be seen to decrease over time while the sum of DE and SE segments decreases slightly. The DE and SE segment numbers at the end of each DS period define the detector configuration used.}
  \label{fig:DS}
\end{figure}

The second improvement introduced here is the division of the dataset into multiple time periods with different segment configurations. This Data Splitting (DS) process better utilizes information from earlier periods with more functional PMTs. 
The DS criteria were
(1) each period contains one full reactor operational cycle, resulting in five DS periods;
(2) all periods have reactor off data bracketing the included reactor on cycle, with the exception of period 1 since the first data collected by PROSPECT was with the reactor on; and
(3) reactor-off data is divided between adjacent periods approximately equally.
For three of the four period divisions, calibration campaigns provide a convenient break point.
All PMTs that were inoperative at any point during a DS period were excluded. Fig.\ref{fig:DS} illustrates the time evolution of DE and SE segments.
A reconstructed energy spectrum of IBD events is then formed for every DS period. The total number of IBD events detected is defined as $\text{N}_{\text{IBD}} = \text{N(E)}_{\text{On}} - \text{R} \cdot \text{N(E)}_{\text{Off}}$, where $\text{N}_{\text{On(Off)}}$ corresponds to the detected IBD-like candidates during reactor-on(off) periods with accidental backgrounds subtracted, $\text{E}$ runs over the energy region [0.8-7.4] MeV, and $\text{R}$ is the relative on-to-off live-time ratio~\cite{PROSPECT:2020sxr}.

With SEER and DS implemented, the IBD selection was optimized to maximize the effective number of signal IBD events ($N_{\rm eff}$) using 20$\%$ of the full dataset, randomly sampled. $N_{\rm eff} \equiv \sum_i ({\rm N}_{\rm IBD,i} / \sigma_i )^2$, where $\sigma_i$ includes the statistical uncertainty of both signal and background and the sum runs over 0.2~MeV-wide prompt energy bins from 0.8 to 7.4~MeV.
$N_{\rm eff}$ illustrates the statistical significance of the measured IBD signal by incorporating the combined statistical uncertainty of both signal and background, being equal to the number of background-free signal events that yield equivalent statistical uncertainty.
Table~\ref{tab:stats} summarizes the relevant event rates for the previous and current analyses. 
The increase in $N_{\rm eff}$ is largely due to the improved background rejection capabilities, as well as an increase in active volume.


\begin{table}[h]
    \begin{tabular}{ccccc}
        \textbf{Data Set} & \textbf{Rx-On(Off) Days} & $\textbf{N}_{\textbf{IBD}}$  & \textbf{ $\mathbf{N_{\rm \mathbf{eff}}}$} & \textbf{S:CB(AB)} \\
        \hline
        \hline
        Prev. Analysis & 95.65(73.09) & 50560 $\pm$ 406 & 18100 & 1.37(1.78) \\ 
        \hline
        This Analysis & 95.62(72.69) & 61029 $\pm$ 338 & 36204 & 3.90(4.31)  \\
        \hline
        Period 1 & 9.54(14.58) & 6357 $\pm$ 100 & 4328 & 4.03(6.21) \\
        Period 2 & 22.83(15.71) & 16546 $\pm$ 172 & 10259 & 4.35(4.64)\\
        Period 3 & 23.20(16.40) & 15094 $\pm$ 166 & 9050 & 4.04(4.44) \\
        Period 4 & 22.29(16.79) & 13486 $\pm$ 161 & 7742 & 3.72(3.39)\\
        Period 5 & 17.76(9.21) & 9546 $\pm$ 146 & 4825 & 3.38(2.88) \\
    \end{tabular}
    \caption{Final IBD event statistics for the previous and current analysis. Reactor-on (RxOn) and -off (RxOff) data taking time is presented in units of calendar days. $N_{\rm IBD}$, $N_{\rm eff}$ and both signal to cosmogenic background (S:CB) and signal to accidental background (S:AB) ratios are calculated over the IBD energy region of [0.8, 7.2]~MeV for previous analysis in \cite{PROSPECT:2020sxr} and [0.8, 7.4]~MeV for the current analysis.}
    \label{tab:stats}
\end{table}

The final analysis method introduced in this Letter is a multi-period detector response unfolding. 
This procedure enables the combination of spectra measured with varying detector responses in the different DS periods into a single antineutrino energy spectrum that can be compared to reactor models or other experimental measurements.
The PROSPECT \uFive{} spectrum analysis uses the WienerSVD approach to perform the unfolding~\cite{Tang:2017rob}. Descriptions of the PROSPECT unfolding approach can be found in the previous joint analyses with the Daya Bay~\cite{DayaBay:2021owf} and STEREO~\cite{Stereo:2021wfd} collaborations.
The key difference in this analysis is that the separate PROSPECT DS periods are treated as correlated rather than uncorrelated inputs to a joint spectrum.
The unfolding process uses the Huber-Mueller \uFive{} model  (HM)~\cite{HM:Huber,HM:Mueller} as the assumed form of the antineutrino spectrum when constructing the Wiener filter.  

The input to the analysis from each DS period includes the corresponding IBD prompt spectrum as well as response and covariance matrices. Each of the five prompt spectra spans a range of prompt energy of [0.8, 7.4]~MeV divided into 33 bins of 0.2~MeV width. Both non-fuel and non-equilibrium contributions from the reactor~\cite{PROSPECT:2020vcl} have been subtracted. 
Response and covariance matrices are generated for each DS period using a well-benchmarked simulation following the procedure in~\cite{PROSPECT:2020sxr}. 
 The five prompt DS spectra are combined into a 165-bin joint energy spectrum vector.
Response and covariance matrices for each DS period are combined into their joint counterparts.
The output of the unfolding framework is a 26-bin antineutrino energy spectrum spanning the energy range of [1.8, 8.3]~MeV, with bin widths of 0.25~MeV. \par
Jointly unfolding data from closely related measurements requires consideration of correlated uncertainties between datasets.
Uncertainties considered as period-correlated are the liquid scintillator energy response, smearing of energy resolution due to liquid scintillator degradation \cite{PROSPECT:2020sxr}, optical grid panel thickness, fiducialization along the length of the cell, and data acquisition thresholds. 
Systematic uncertainties from background variations and IBD spectrum background subtraction are considered to be uncorrelated between periods.  
The energy bin and period uncertainty correlations for each effect are included in a joint covariance matrix as on- and off-diagonal blocks which are produced through the generation and analysis of systematically fluctuated MC datasets~\cite{PROSPECT:2020sxr}.  \par

\begin{figure}[t]
  \centering
  \includegraphics[width=0.48\textwidth]{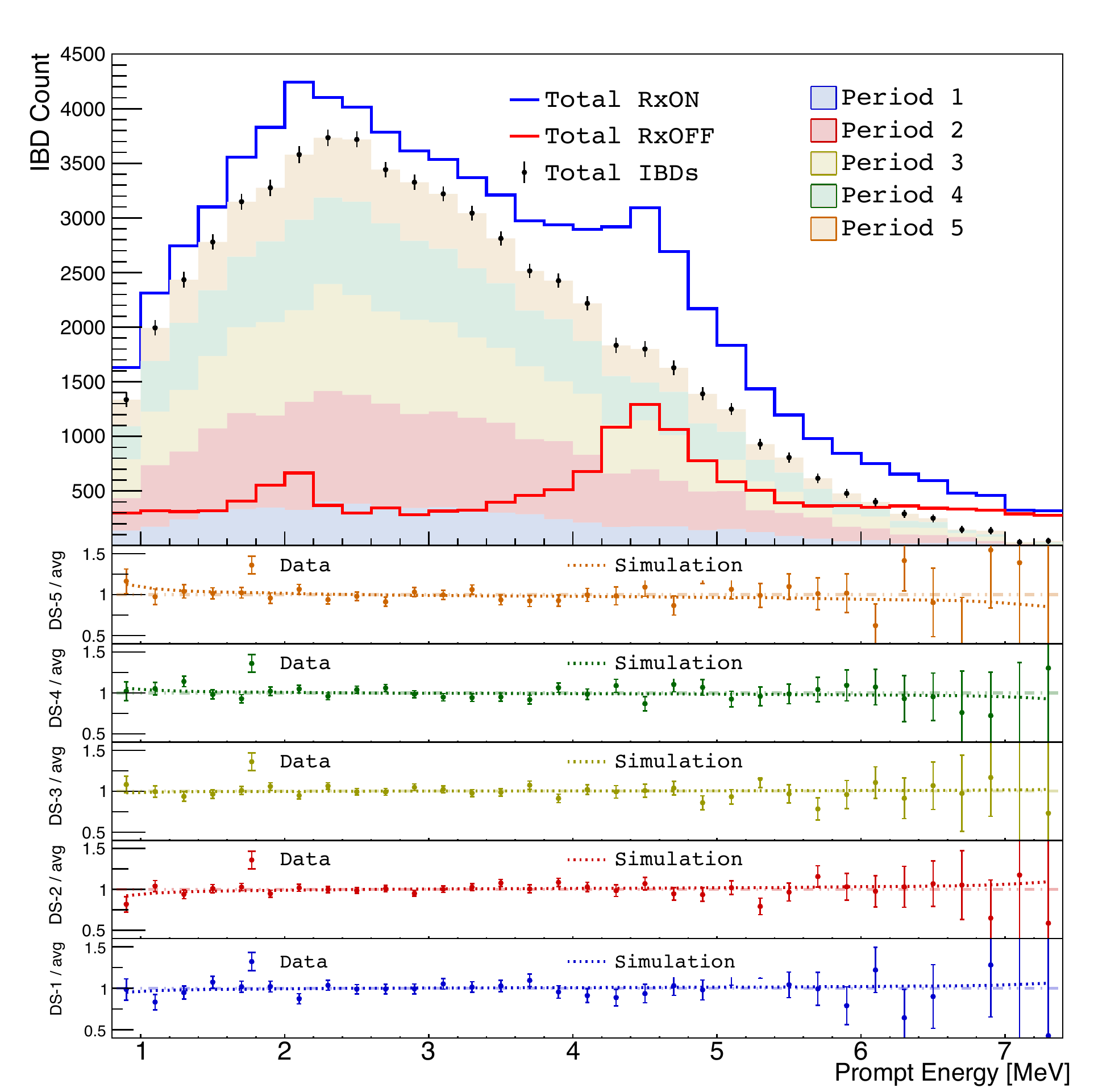}
  \caption{(first panel) Solid blue and red lines represent total reactor-on (RxOn) and -off (RxOff) prompt spectra respectively for all periods, while the point distribution is the result of their subtraction. Colored areas represent measured IBD prompt spectra for each period. (second to sixth panels) Point distributions represent the ratio between periods 1-5 and the total spectrum normalized to the integrated counts of each period, while the dotted line displays the expected behavior according to Monte Carlo simulations of the different detector configurations.}
  \label{fig:prompt}
\end{figure}

Fig.~\ref{fig:prompt} depicts the aggregated sum of all prompt spectra. 
The compatibility between periods is demonstrated with the lower panels in Fig.~\ref{fig:prompt}, which display the ratio of each period to the average spectrum. The latter is calculated as the period-normalized sum of all prompt spectra. 
The dotted lines are constructed similarly to the point distributions, using HM spectrum folded through each period's response matrix.  
Despite substantial differences in inoperative PMT channel counts, detector response is comparable between DS periods, as indicated by the flatness of these ratios. Slight MC-predicted response differences are present as minor deviations at low and high energy edges. Measurement compatibility between periods has also been confirmed by folding individual periods' data into matching reconstructed energy spaces, using the formalism described in \cite{DayaBay:2021owf, Stereo:2021wfd}.  

\begin{figure}[t]
  \centering
  \includegraphics[width=0.48\textwidth]{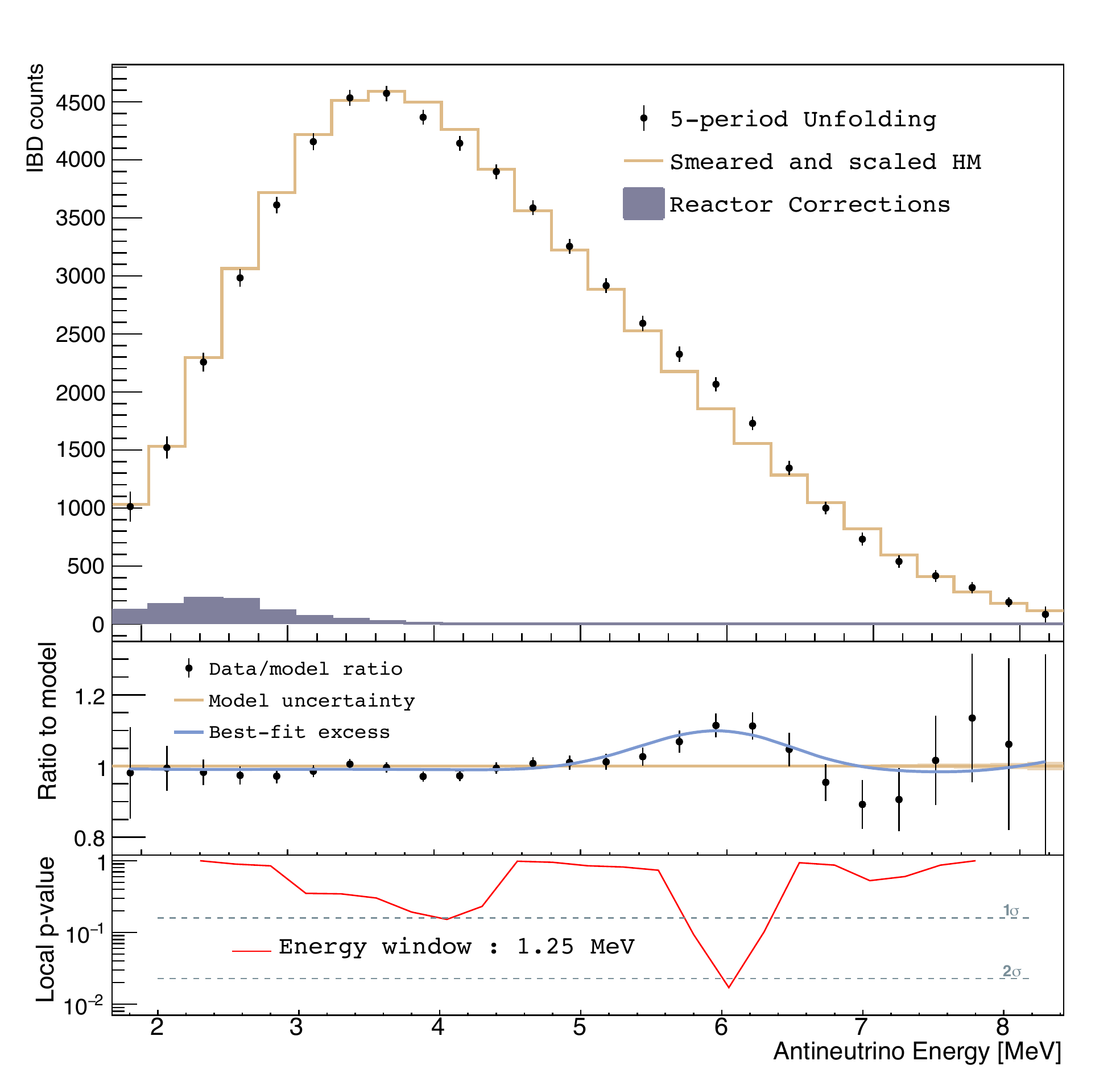}
  \caption{Unfolded antineutrino spectrum from HEU reactor at HFIR. (first panel) Points represent PROSPECT's unfolded antineutrino spectrum; the magenta line depicts HM \uFive~prediction; and the gray shaded histogram corresponds to the reactor non-fuel and non-equilibrium contributions, which have been already subtracted from the data. The transformed covariance matrix is also displayed. (second panel) Data to model ratio together with the Gaussian function $G_{\text{PRO}}(A\cdot r, \mu, \sigma)$ obtained from the simultaneous fit of PROSPECT's and DYB's unfolded spectra. (third panel) The local p-value obtained from a 1.25~MeV-wide sliding windows analysis.}
  \label{fig:unfold}
\end{figure}

The 5 prompt spectra from Fig.~\ref{fig:prompt} are simultaneously unfolded into neutrino space by using the WienerSVD framework. The resulting unfolded antineutrino spectrum is compared to HM (normalized to the integral of all periods) in Fig.~\ref{fig:unfold}. To account for unfolding bias, the smearing matrix obtained from using the WienerSVD method over the data has also been applied to the model.


Agreement between data and model can be quantified by using the covariance matrix formalism described in \cite{PROSPECT:2020sxr}, 
\begin{equation}\label{eq:chi2}
    \bar{\chi}^2 = [S-M]^T C^{-1} [S-M] / {\rm ndf}
\end{equation} 
\noindent where $S$ and $M$ represent signal and signal-normalized model prediction respectively, ${\rm ndf}$ refers to the number of degrees of freedom, and $C = C_{\text{PRO}} + C_{\text{HM}}$ corresponds to the sum of their respective covariance matrices. Applying Eq. \ref{eq:chi2} shows general agreement between the distributions, with a $\chi^2$ over the number of degrees of freedom of $\bar{\chi}_{\text{HM}}^2 = 30.2/25$, with a single-sided p-value of $p = 0.22$. A local discrepancy at energies above 5~MeV can be observed in Fig. \ref{fig:unfold}. This can be quantified by means of a sliding window method \cite{DayaBay:2017, PROSPECT:2018snc}, where a set of nuisance parameters are added in 1.25~MeV-wide windows along the spectrum, guided by the scale of deviations observed in previous experiments. Then, the $\Delta\chi^2$ with respect to the best fit and its corresponding single-sided p-value are determined for each window as depicted in Fig~\ref{fig:unfold}, confirming the local excess.

As discussed earlier, such an excess of antineutrinos in the 5-7~MeV region has been consistently observed in modern reactor antineutrino experiments. One case study is Daya Bay's (DYB) LEU-derived \uFive{} spectra obtained from a time-evolution study on the fission fractions at the Daya Bay nuclear power complex ~\cite{DayaBay:2021dqj}. When comparing PROSPECT’s full HEU- and DYB's LEU-derived \uFive{} spectra through Eq.~\ref{eq:chi2} one obtains $\bar{\chi}_{\text{DYB}}^2 = 20.2/23$, with $p = 0.63$, indicating better consistency between both experimental \uFive{} spectra than with HM.

The isotopic origin of the excess can be probed by contrasting its appearance in composite HEU- and LEU-based measurements~\cite{hayes:bump,buck:bump,Huber:2016xis,PROSPECT:2018snc, PROSPECT:2020sxr}.

\begin{figure}[t]
  \centering
    \includegraphics[width=0.48\textwidth]{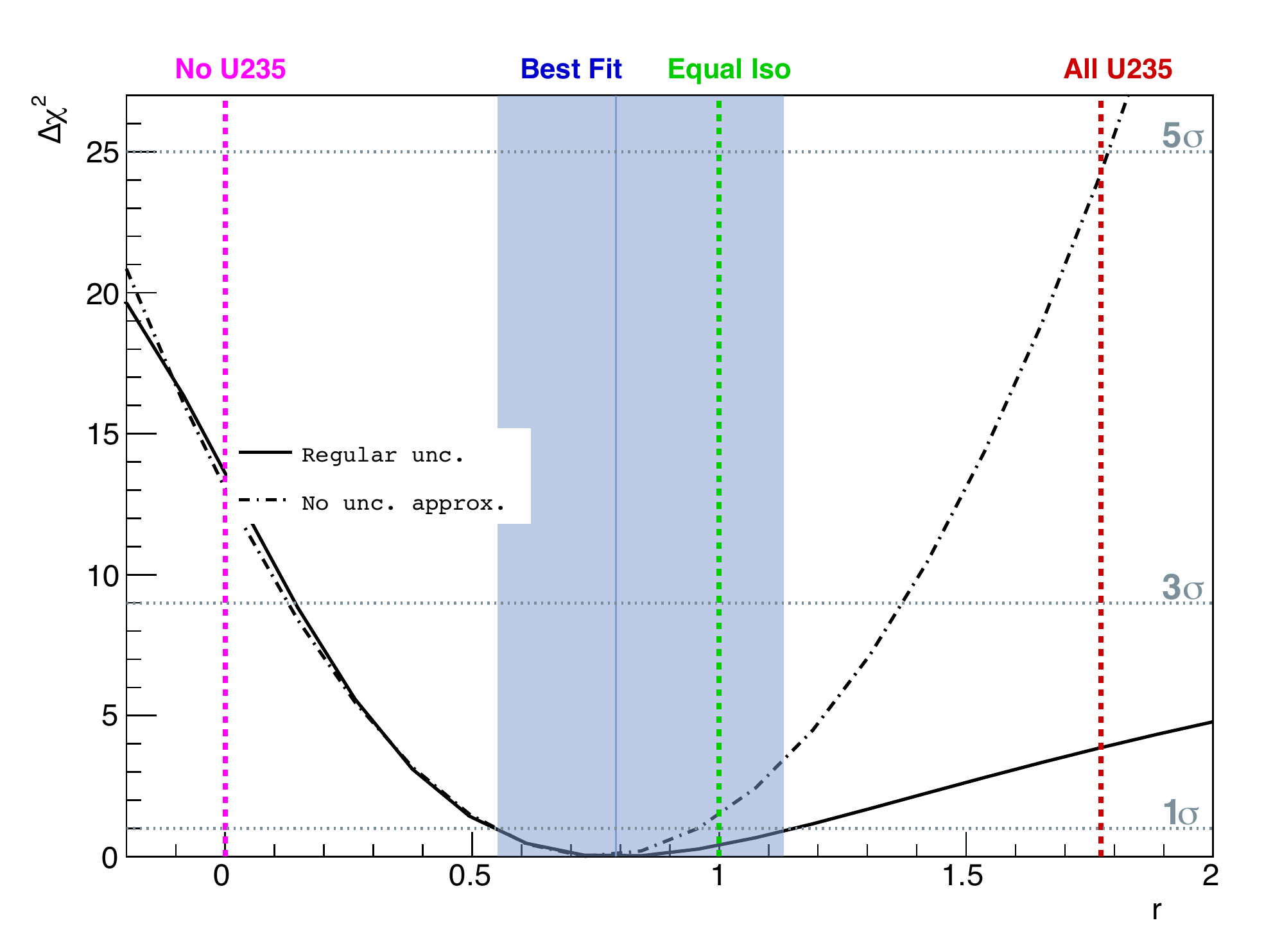}
    \caption{$\Delta \chi^2$ distributions for a Gaussian-like excess simultaneous fit of PROSPECT's and DYB's unfolded spectra. Profile from the common fit (solid-black) displays an asymmetric behavior due to the negligible impact of DYB's spectrum in rejecting the No-\uFive{} hypothesis. The best-fit value of $r = 0.79^{+0.35}_{-0.25}$ is found with a $\Delta \chi^2 = \DXNull(\DXAll)$ with respect to $r = 0(1.77)$, amplitude ratio required to satisfy No-\uFive{} (All-\uFive) hypothesis and represented by the magenta (red) line. The green dotted line represents the equal-isotope hypothesis, i.e. $r = 1$, which crosses the profile at $\Delta \chi^2 = \DXIso$. When DYB's uncertainties are largely down-scaled (dotted-black), $\Delta \chi^2$ becomes symmetric.}
  \label{fig:DeltaChi2}
\end{figure}

For this study, the excess in DYB's multi-isotope unfolded LEU spectrum~\cite{DayaBay:2021dqj} has been used as the LEU-based case study to contrast with PROSPECT's HEU-based measurements. The magnitude of the excess for each of the experiments is obtained by comparing the unfolded spectra with altered versions of their corresponding fuel-modeled HM spectra. Such alteration comes as a Gaussian function, $G_{\text{PRO}}(A\cdot r, \mu, \sigma)$ or $G_{\text{DYB}}(A, \mu, \sigma)$, added to each of the models respectively. Both functions are introduced with matching mean $\mu$ and standard deviation $\sigma$, while their amplitudes are correlated by the parameter $r$. The data-model comparison is performed for both experiments simultaneously in a common fit of parameters, resulting in an excess amplitude of $A = 11 \pm 4 \%$ and an excess ratio of $r = 0.79^{+0.35}_{-0.25}$, with a $\bar{\chi}^2_{\text{fit}} = 34.8/44$ ($p = 0.84$). The resulting excess is shown in the second panel of Fig.~\ref{fig:unfold}. The $\Delta\chi^2$ distribution in terms of $r$ is represented by the solid line in Fig~\ref{fig:DeltaChi2}.
Given the excess amplitude obtained for pure \uFive{} and an LEU spectrum, quantitative statements can be made about the presence or absence of spectral mis-modelling for different fission isotopes. Three cases of interest are commonly discussed in the literature: (1) \uFive{} has no observed excess, and instead, sub-dominant fission isotopes bear responsibility for data-model disagreements (No-\uFive{} hypothesis), (2) all isotopes are equal contributors (equal-isotope hypothesis), and (3) \uFive{} is solely responsible for all observed excesses (All-\uFive{} hypothesis).
In the No-\uFive~case, a null excess condition $r = 0$ would need to be satisfied. The No-\uFive{} case profiles in Fig.\ref{fig:DeltaChi2} at $\Delta \chi^2 = \DXNull$ ($p = 2.26\cdot10^{-4}$), indicating that PROSPECT data disfavors this hypothesis at $\CLNull{}\sigma$. In the equal-isotope hypothesis case, PROSPECT and DYB should observe identically-sized excesses, i.e $r = 1.0$. In Fig. \ref{fig:DeltaChi2}, this case profiles at $\Delta \chi^2 = \DXIso$ ($p = 0.53$), from which it can be concluded that PROSPECT and DYB data are consistent with all isotopes contributing equally to the excess. Finally, under the All-\uFive{} hypothesis, the excess observed in the composite spectrum of DYB would be produced only by the average effective fraction of fissions undergone by \uFive{} of 0.564~\cite{DayaBay:2021dqj}.  Thus, this hypothesis requires PROSPECT to observe an excess amplitude larger than DYB's, i.e $r = 1.77.$  This point profiles at $\Delta \chi^2 = \DXAll$ ($p = 0.05$) in Fig.~\ref{fig:DeltaChi2}, which indicates that PROSPECT's data disfavors the All-\uFive{} hypothesis at $\CLAll{}\sigma$.
Uncertainties in DYB's excess amplitude, which are dominated by detector systematics that were not previously considered in PROSPECT analyses, have a major effect in the hypothesis rejection. This effect is illustrated by the dashed profile in Fig.~\ref{fig:DeltaChi2} which results from down-scaling DYB's covariance matrix by four orders of magnitude, rendering DYB's uncertainties negligible for the fit.
In such a scenario, the hypothesis rejection power from the fit increases to almost $5\sigma$ for All-\uFive, notably improving over results presented by PROSPECT in \cite{PROSPECT:2020sxr}.
Given the substantial role of uncorrelated detector systematics in this comparison, future measurements of HEU and LEU cores with a common detector~\cite{Gebre:2017vmm, PROSPECT:2021jey, Akindele:2022dpr} would be particularly valuable in further improving tests of the All-\uFive{} hypothesis.

In summary, a new analysis of PROSPECT's complete reactor antineutrino data set has been performed. This effort incorporates new analysis techniques including Single Ended Event Reconstruction and Data Splitting, resulting in a multi-period analysis with significantly improved antineutrino event selection criteria.
The recovery of more than $10000$ IBD candidates and a major improvement in background rejection has nearly doubled PROSPECT's statistical power and increased the average signal-to-background ratio from 1.58 to 4.11.  
Using this improved dataset, one of the most precise measurements of the \uFive{} antineutrino energy spectrum to date is obtained from the HFIR HEU reactor via a simultaneous multi-period unfolding through the WienerSVD formalism.  
Comparisons of measured spectra with Huber-Mueller \uFive{} model and LEU-based measurements show consistent high-energy spectral excesses between HEU and LEU reactors, disfavoring both No-\uFive{}  and All-\uFive{} hypotheses at \CLNull{}$\sigma$ and \CLAll{}$\sigma$ respectively as explanations for the 5-7~MeV excess. In consonance with recent modeling advancements and fission beta measurements \cite{fallot2, kopeikin}, these final PROSPECT-I results further reinforce the theoretically plausible scenario that all primary fission isotopes suffer from incorrectly predicted reactor antineutrino spectra.


This material is based upon work supported by the following sources: US Department of Energy (DOE) Office of Science, Office of High Energy Physics under Award No. DE-SC0016357 and DE-SC0017660 to Yale University, under Award No. DE-SC0017815 to Drexel University, under Award No. DE-SC0008347 to Illinois Institute of Technology, under Award No. DE-SC0016060 to Temple University, under Award No. DE-SC0010504 to University of Hawaii, under Contract No. DE-SC0012704 to Brookhaven National Laboratory, and under Work Proposal Number  SCW1504 to Lawrence Livermore National Laboratory. This work was performed under the auspices of the U.S. Department of Energy by Lawrence Livermore National Laboratory under Contract DE-AC52-07NA27344 and by Oak Ridge National Laboratory under Contract DE-AC05-00OR22725. Additional funding for the experiment was provided by the Heising-Simons Foundation under Award No. \#2016-117 to Yale University.
 
J.G. is supported through the NSF Graduate Research Fellowship Program. This work was also supported by the Canada First Research Excellence Fund (CFREF), and the Natural Sciences and Engineering Research Council of Canada (NSERC) Discovery  program under grant \#RGPIN-418579, and Province of Ontario.
 
We further acknowledge support from Yale University, the Illinois Institute of Technology, Temple University, University of Hawaii, Brookhaven National Laboratory, the Lawrence Livermore National Laboratory LDRD program, the National Institute of Standards and Technology, and Oak Ridge National Laboratory. We gratefully acknowledge the support and hospitality of the High Flux Isotope Reactor and Oak Ridge National Laboratory, managed by UT-Battelle for the U.S. Department of Energy.

\bibliography{references}

\end{document}